\documentclass[superscriptaddress,showpacs,amssymb,10pt,aps,prl,longbibliography,reprint]{revtex4-1}

\usepackage{graphicx,epsfig,amssymb,times} 
\usepackage{amsmath,amsfonts}
\usepackage{bm}
\usepackage{epstopdf}
\usepackage[linktocpage,colorlinks]{hyperref}
\usepackage[caption=false]{subfig}
\usepackage[usenames]{color}

\definecolor{coolblack}{rgb}{0.0, 0.18, 0.39}
\definecolor{darkred}{rgb}{0.5,0,0}
\definecolor{darkgreen}{rgb}{0,0.5,0}
\definecolor{darkblue}{rgb}{0,0,0.5}
\definecolor{lapislazuli}{rgb}{0.15, 0.38, 0.61}
\definecolor{venetianred}{rgb}{0.78, 0.03, 0.08}
\definecolor{bleudefrance}{rgb}{0.19, 0.55, 0.91}
\definecolor{dogwoodrose}{rgb}{0.84, 0.09, 0.41}
\hypersetup{colorlinks=true, citecolor=darkblue, linkcolor=darkblue, 
urlcolor = darkblue}
\def\be{\begin{equation}}
\def\ee{\end{equation}}

\newcommand{\bea}{\begin{eqnarray}}
\newcommand{\eea}{\end{eqnarray}}
\newcommand{\ben}{\begin{enumerate}}
\newcommand{\een}{\end{enumerate}}
\newcommand{\bi}{\begin{itemize}}
\newcommand{\ei}{\end{itemize}}

\renewcommand{\d}{\partial}

\def\ga{\mathrel{\raise.3ex\hbox{$>$\kern-.75em\lower1ex\hbox{$\sim$}}}}
\def\la{\mathrel{\raise.3ex\hbox{$<$\kern-.75em\lower1ex\hbox{$\sim$}}}}

\def\l{\left}
\def\r{\right}
\def\be{\begin{equation}}
\def\ee{\end{equation}}

\renewcommand{\d}{\rm{d}}

\def\I_M{{I_{\scriptscriptstyle M\times M}}}

\def\be{\begin{equation}}
\def\ee{\end{equation}}
\def\bea{\begin{eqnarray}}
\def\eea{\end{eqnarray}}
\newcommand{\beq}{\begin{eqnarray}}
\newcommand{\eeq}{\end{eqnarray}}

\newcommand{\beqal}{\begin{eqnarray}\label}
\newcommand{\beqa}{\begin{eqnarray}}
\newcommand{\eeqa}{\end{eqnarray}}

\newcommand {\non}{\nonumber\\}
\newcommand {\f}{\frac}
\newcommand {\lmw}{lm\omega}

\newcommand{\sigabs}{\sigma_{\text{abs}}}

\newcommand{\sigraygeo}{\sigma_{\text{geo}}}    % {\sigma^{\text{ray}}_{\text{geo}}}

\newcommand{\sw}{\mathfrak{s}}  % spin-weight
\newcommand{\calA}{\mathcal{A}}
\newcommand{\calB}{\mathcal{B}}
\newcommand{\calC}{\mathcal{C}}
\newcommand{\eps}{\varepsilon}

\begin{document}
\title{\large Absorption of electromagnetic and gravitational waves by Kerr black holes: \\ Shadows, superradiance and the spin-helicity effect }

\author{Luiz C. S. Leite}
\email{lcleite1@sheffield.ac.uk}
\affiliation{Consortium for Fundamental Physics, School of 
Mathematics and Statistics, University of Sheffield, Hicks Building,
Hounsfield Road, Sheffield S3 7RH, United Kingdom}
\affiliation{Faculdade de F\'{\i}sica, Universidade 
Federal do Par\'a, 66075-110, Bel\'em, Par\'a, Brazil.}
\author{Sam R.~Dolan}
\email{s.dolan@sheffield.ac.uk}
\affiliation{Consortium for Fundamental Physics, School of 
Mathematics and Statistics, University of Sheffield, Hicks Building,
Hounsfield Road, Sheffield S3 7RH, United Kingdom}
\author{Lu\'is C. B. Crispino}
\email{crispino@ufpa.br}
\affiliation{Faculdade de F\'{\i}sica, Universidade 
Federal do Par\'a, 66075-110, Bel\'em, Par\'a, Brazil.}
\begin{abstract}
We study the absorption of plane 
waves by Kerr black holes. We calculate the absorption cross section: the area of the black hole shadow at a finite wavelength. We present a unified picture of the absorption of all massless bosonic fields, focussing on the on-axis incidence case. We investigate the spin-helicity effect, arising from a coupling between dragging of frames and the helicity of a polarized wave. We introduce and calibrate an extended sinc approximation which provides new quantitative data on the spin-helicity effect in strong-field gravity. 
\end{abstract}

\date{\today}
\maketitle

{\it Introduction.}---Black holes, once dismissed as a mathematical artifact of Einstein's theory of general relativity (GR), have come to play a central role in modern astronomy and theoretical physics \cite{Thorne:1986, Frolov:1998wf}. In astronomy, black holes provide a solution: in galaxy formation scenarios, in active galactic nuclei and in core-collapse supernovae, for instance. In theoretical physics, black holes pose a challenge: as spacetime curvature grows without bound in GR, the classical theory breaks down. Yet, novel quantum gravity effects apparently remain shrouded by a horizon endowed with generic thermodynamic properties \cite{Wald:1999vt}. 

Two recent advances in interferometry have opened new data channels on astrophysical black holes. In September 2015, LIGO detected the first gravitational-wave signal: a characteristic `chirp' from a black hole binary merger \cite{Abbott:2016blz}. Hundreds more chirps are anticipated over the next decade \cite{TheLIGOScientific:2016pea}. In April 2017, the Event Horizon Telescope (EHT) \cite{Ricarte:2014nca} -- a global array of radio telescopes linked by very long baseline interferometry -- observed the supermassive black hole candidates Sgr.~A* and M87* at a resolution three orders of magnitude beyond that of the Hubble telescope \cite{castelvecchi2017hunt}. Ultimately, the EHT will seek to study the black hole shadow itself \cite{Falcke:1999pj, Lu:2014zja, Johannsen:2016uoh}, using techniques to surpass the diffraction limit \cite{Akiyama:2017rcc}. 

These experimental advances motivate study of the interaction of electromagnetic waves (EWs) and gravitational waves (GWs) with black holes \cite{Glampedakis:2001cx, Dolan:2008kf, Caio:2013}. EWs and GWs propagating on curved spacetimes in vacuum share some traits. For example, both possess two independent (transverse) polarizations that are parallel-transported along null geodesics in the ray-optics limit. 
Yet there are key physical differences. GWs are tenuous, in the sense that they are not significantly attenuated or rescattered by matter sources. GWs are typically long-wavelength and polarized, because rotating quadrupoles (for example, binary systems or asymmetric neutron stars) predominantly emit circular-polarized waves at twice the rotational frequency \cite{Abbott:2016bqf}. For example, $\lambda \sim 10^{-3} \text{m}$ for EHT observations, whereas $\lambda \sim 10^7 \text{m}$ for GW150914.

In this Letter we examine the absorption of a monochromatic planar wave of frequency $\omega$ incident upon a Kerr black hole of mass $M$ and angular momentum $J$ in vacuum. We calculate the absorption cross section $\sigabs$, i.e., the cross-sectional area of the black hole shadow \cite{Falcke:1999pj, Lu:2014zja, Johannsen:2016uoh} beyond the ray-optics approximation. For the first time, we present unifying results for scalar ($s=0$), electromagnetic ($s=1$) and gravitational ($s=2$) waves. Our results highlight the influence of two key phenomena: superradiance and the spin-helicity effect, described below.

\begin{figure}
 \begin{center}
  \includegraphics[height=6.5cm]{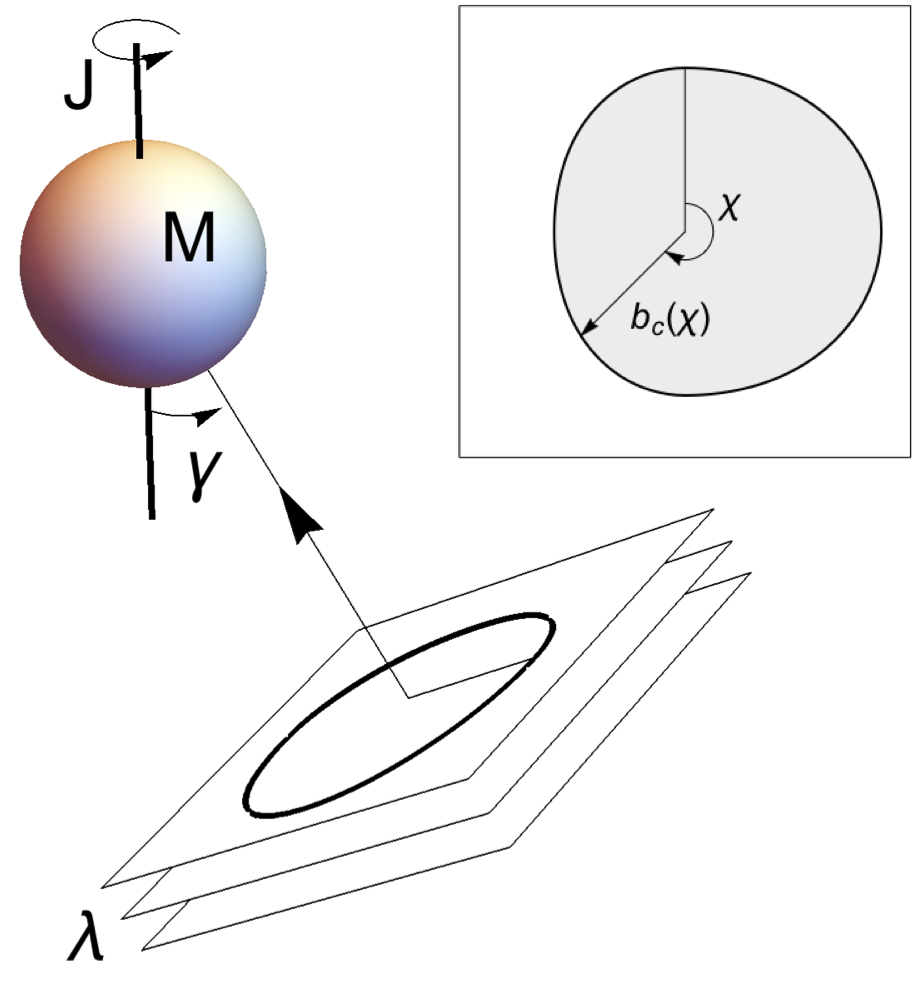}
 \end{center}
 \caption{A planar wave of frequency $\omega = 2\pi c / \lambda$ incident upon a rotating black hole of mass $M$ and angular momentum $J$ at an angle $\gamma$. \emph{Inset:} the locus $b_c(\chi)$ of the black hole shadow on the wavefront.
 }
 \label{fig:sketch}
\end{figure}

The absorption scenario, illustrated in Fig.~\ref{fig:sketch}, is encapsulated by several dimensionless parameters: the ratio of the gravitational length to the (reduced) wavelength $G M \omega / c^3$; the dimensionless black hole spin $a^\ast \equiv a / M$ ($0 \le a^\ast < 1$) where $a = J c^2 / G M$; the spin of the field $s = 0, 1, 2$; the angle of incidence with respect to the black hole axis $\gamma$; and the helicity of the circular polarization $\pm 1$. We adopt natural units such that $G = c = 1$.

{\it Black hole shadows.}--- In the geometric-optics limit ($\lambda \rightarrow 0$), an observer studying a black hole in vacuum with a pinhole camera will see a dark region on the image plane defined by the set of null-geodesic rays entering the pinhole which, when traced backwards in time, pass into the black hole (BH). The boundary of the shadow is determined by those rays which asymptote towards an (unstable) photon orbit. In Schwarzschild spacetime, an observer at radial coordinate $r_0$ sees a shadow of angular radius $\alpha$ where \cite{Synge:1966}
 \beq
 \sin^2 \alpha = \frac{27}{4} \frac{(\rho - 1)}{\rho^3} , \quad \quad \rho \equiv \frac{r_0 c^2}{G M}.
 \eeq
 For Sgr A*, $\alpha \approx 25 \,\mu \text{arcsec}$, with $r_0 \approx 8.3 \,\text{kpc}$ and $M \approx 4.1 \times 10^6 M_\odot$ \cite{Grenzebach:2014fha}. In Kerr spacetime, $\alpha$ is a function of angle $\chi$ relative to the (projected) spin axis. Alternatively, the shadow can be defined via rays orthogonal to a planar surface, as shown in Fig.~\ref{fig:sketch}. Far from the black hole, the impact parameter defining the shadow boundary is $b(\chi) = r_0 \alpha(\chi) + O(\rho^{-1})$, and the cross section is 
$
 \sigraygeo = \tfrac{1}{2} \int_0^{2\pi} b^2_c(\chi) d\chi .  \label{eq:sigray}
$

{\it Superradiance and spin-helicity.}--- 
Superradiance is a radiation-enhancement mechanism by which a black hole may shed mass and angular momentum and yet still increase its horizon area, and thus its Bekenstein-Hawking entropy \cite{Brito:2015oca}. As a consequence, $\sigabs$ may become negative at low frequencies, through stimulated emission. The effect is strongly enhanced by spin $s$. 

The spin-helicity effect is a coupling between a rotating source, such as a Kerr black hole, and the helicity of a polarized wave of finite wavelength $\lambda$ \cite{Mashhoon:1974cq}. A rotating spacetime distinguishes and separates waves of opposite helicity \cite{Frolov:2011mh, Frolov:2012zn, Yoo:2012vv}. In the weak-field, rays are deflected through an angle $\zeta \Theta_E$, with $\Theta_E \equiv \frac{4GM}{c^2b}$ the Einstein angle and $\zeta = 1 + \ldots$ an asymptotic series in which the spin-helicity effect is anticipated at $O\left( \frac{J \lambda}{M c b^2} \right)$ \cite{Mashhoon:1974cq}. In the strong-field, we anticipate that waves with a counter-rotating circular polarization are preferentially absorbed ($\sigabs^- > \sigabs^+$). 

%%%%%%%%%%%%%%%%%%%%%%%%%%%%%%%%%%%%%%%%%%%%%%%%%%%%%%
{\it Wave propagation on the Kerr spacetime.}---
The Kerr spacetime is described in Boyer-Lindquist coordinates $\{t,r,\theta,\phi\}$ by the line element
\bea
{\d} s^2 &=& -\f{1}{\Sigma}\l(\Sigma-2Mr\r) dt^2
 - \frac{4Mar\sin^2\theta}{\Sigma} dt d\phi 
 + \frac{\Sigma}{\Delta} dr^2 
\nonumber\\
&& 
+ \Sigma d\theta^2
+ \frac{(r^2+a^2)^2\sin^2\theta-\Delta 
a^2\sin^4\theta}{\Sigma} d\phi^2,\label{eq:linelement}
\eea
where $\Sigma\equiv r^2+a^2\cos^2\theta$, and $\Delta\equiv r^2-2Mr+a^2$. We focus on the $a^2 < M^2$ case of a rotating BH with two distinct horizons: an internal~(Cauchy) horizon located at~$r_{-}=M-\sqrt{M^2-a^2}$ and 
an external~(event) horizon at~$r_{+}=M+\sqrt{M^2-a^2}$.

In the vicinity of a Kerr black hole, perturbing fields are described by a single master equation, first obtained by Teukolsky \cite{teukolsky1972rotating} using the Newman-Penrose formalism. In vacuum the 
master equation takes the form
\bea
\left[\frac{(r^{2}+a^{2})^{2}}{\Delta}-a^{2}\sin^{2}\theta\right]\frac{\partial^{2}\psi}{\partial t^{2}}+\frac{4Mar}{\Delta}\frac{\partial^{2}\psi}{\partial t\partial\phi}\non+\left[\frac{a^{2}}{\Delta}-\frac{1}{\sin^{2}\theta}\right]\frac{\partial^{2}\psi}{\partial\phi^{2}}
-\Delta^{-\sw}\frac{\partial}{\partial r}\left(\Delta^{\sw+1}\frac{\partial\phi}{\partial r}\right)\non-\frac{1}{\sin\theta}\frac{\partial}{\partial\theta}\left(\sin\theta\frac{\partial\psi}{\partial\theta}\right)+(\sw^{2}\cot^{2}\theta-\sw)\psi\nonumber \\
-2\sw\left[\frac{a(r-M)}{\Delta}+\frac{i\cos\theta}{\sin^{2}\theta}\right]\frac{\partial\psi}{\partial\phi}\non-2\sw\left[\frac{M(r^{2}-a^{2})}{\Delta}-r-ia\cos\theta\right]\frac{\partial\psi}{\partial t}=0,\label{eq:master_eq}
\eea
where $\sw$ is the spin-weight of the field. We use $\sw = -s$ throughout, where $s = 0, 1, 2$ for scalar, electromagnetic and gravitational fields, respectively. 
One can separate variables in Eq.~\eqref{eq:master_eq} using the standard ansatz
\be
\psi_{\sw\lmw}(t,\,r,\,\theta,\,\phi)=R_{\sw\lmw}(r)S_{\sw\lmw}(\theta)e^{-i(\omega t-m\phi)},\label{eq:ansatz}
\ee
to obtain angular and radial equations, 
\bea
\frac{1}{\sin\theta}\frac{d}{d\theta}\left(\sin\theta\frac{dS_{\sw\lmw}}{d\theta}\right) + U_{\sw\lmw}(\theta) S_{\sw\lmw} &=& 0 , \label{eq:angular_eq} \\
\Delta^{-\sw}\frac{d}{dr}\l(\Delta^{\sw+1}\frac{dR_{\sw\lmw}}{dr}\r)+V_{\sw\lmw}(r)R_{\sw\lmw} &=& 0,
\label{eq:radial_eq} 
\eea
where
\bea
U_{\sw\lmw} &\equiv& \lambda_{\sw\lmw} + 2 a m \omega - 2a\omega \sw\cos\theta-\frac{\l(m+a\cos\theta\r)^2}{\sin^{2}\theta}+\sw , \nonumber \\
V_{\sw\lmw} &\equiv& \frac{1}{\Delta}\left[K^{2}-2(r-M)K\right]-\lambda_{\sw\lmw}+4i\omega \sw r ,
\eea
and $K\equiv(r^{2}+a^{2})\omega-am$. The angular functions $S_{\sw\lmw}(\theta)$ are known as 
spin-weighted spheroidal harmonics, and have as limiting cases the spheroidal harmonics ($\sw=0$) and 
the spin-weighted spherical harmonics ($a\omega=0$).

We seek solutions of Eq.~\eqref{eq:radial_eq} that are purely ingoing 
at the event horizon, satisfying the following boundary conditions:
\be
R_{\sw\lmw}\sim\begin{cases}
\mathcal{T}_{\sw\lmw}e^{-\imath(\omega - m\Omega_h) r_{\star}}\Delta^{-\sw} , & r\rightarrow r_+,\\
\mathcal{I}_{\sw\lmw}r^{-1}e^{-\imath\omega r_{\star}} + & \\ \quad \mathcal{R}_{\sw\lmw}r^{-(2\sw+1)}e^{\imath\omega r_{\star}}, & r\rightarrow 
+\infty,\label{eq:ingoing_sol}
\end{cases} 
\ee
where $\Omega_h \equiv \frac{a}{2Mr_+}$ is the angular frequency of the black hole horizon. Here $r_{\star}$ is the tortoise coordinate $r_{\star}\equiv\int \f{(r^2+a^2)}{\Delta} dr$ such that $r_{\star}\rightarrow+\infty$ when $r\rightarrow+\infty$ and $r_{\star}\rightarrow-\infty$ when $r\rightarrow r_+$. 

%%%%%%%%%%%%%%%%%%%%%%%%%%%%%%%%%%%%%%%%%%%%%%%%%%%%%%%%%%%%%%
{\it The absorption cross section.}---For an asymptotic incident plane wave travelling in the direction~$\hat{n}=\sin\gamma \, \hat{x}+\cos\gamma \, \hat{z}$ the absorption cross section $\sigabs$ is given by \cite{Futterman:1988ni}
\be
\sigabs = \frac{4\pi^{2}}{\omega^{2}}\sum_{l=|\sw|}^{+\infty}\sum_{m=-l}^{+l}\left|
S_{\sw\lmw}(\gamma)\right|^{2}\Gamma_{\sw\lmw}. \label{eq:absorption_cs}
\ee
The transmission factor $\Gamma_{\sw\lmw}$ is the ratio of the energy passing into to the hole to that encroaching from infinity, $\frac{dE_{\text{hole}}}{dE_{\text{in}}}$ \cite{Brito:2015oca}. It takes the same sign as $\omega (\omega - m\Omega_h)$, so it is \emph{negative} for low-frequency co-rotating modes. 
Using energy balance, $dE_{\text{hole}} = dE_{\text{in}} - dE_{\text{out}}$, one obtains
\begin{subequations}
\begin{align}
\Gamma_{0\lmw}&=1-\left|\f{\mathcal{R}_{0lm\omega}}{ 
\mathcal{I}_{0lm\omega}}\right|^{2},  \label{eq:transs0} \\
\Gamma_{-1\lmw}&= 1-\f{B^{2}_{\lmw}}{16\omega^4}\left|\f{\mathcal{R}_{-1lm\omega}}{ 
\mathcal{I}_{-1\lmw}}\right|^{2}, \label{eq:transs1} \\
\Gamma_{-2\lmw}&= 1-\f{\text{Re}^2(C)+144M^2\omega^2}{256\omega^8}\left|\f{\mathcal{R}_{-2lm\omega}}{ 
\mathcal{I}_{-2\lmw}}\right|^{2}, \label{eq:transs2}
\end{align}
\label{eq:trans}
\end{subequations}
for the scalar~($\sw=0$), electromagnetic~($\sw=-1$), and gravitational~($\sw=-2$) cases, respectively. Here $B_{\lmw}^{2}\equiv \lambda_{-1\lmw}^2+4am\omega-4a^2\omega^2$, $\text{Re}^2(C)=[(\lambda_{-2\lmw}+2)^2+4am\omega-4(a\omega)^2](\lambda^2_{-2\lmw}+36am\omega-36a^2\omega^2)+(2\lambda_{-2\lmw}+3)(96a^2\omega^2-48am\omega)-144a^2\omega^2$, and $\mathcal{I}_{\sw\lmw}$, $\mathcal{R}_{\sw\lmw}$ are the coefficients appearing in the ingoing solutions of Eq.~\eqref{eq:ingoing_sol}. 

%%%%%%%%%%%%%%%%%%%%%%%%%%%%%%%%%%%%%%%%%%%%%%%%%%%%%%%%%%%%%
{\it Numerical method.}--- In order to determine the absorption cross section via Eq.~\eqref{eq:absorption_cs} we first computed the spin-weighted spheroidal harmonics $S_{\sw\lmw}$ and the transmission factors 
$\Gamma_{\sw\lmw}$ by solving Eqs.~\eqref{eq:angular_eq} 
and~\eqref{eq:radial_eq} with numerical methods. 

We obtained the spin-weighted spheroidal harmonics $S_{\sw\lmw}$ and its corresponding eigenvalues $\lambda_{\sw\lmw}$ using the {\it{spectral eigenvalue method}} as described in Refs.~\cite{Dolan:2008kf, Cook}. We have tested the angular eigenvalues $\lambda_{\sw\lmw}$ obtained via the spectral eigenvalue method against the low-$a\omega$ formula provided in Ref.~\cite{berti2006eigenvalues}, obtaining a satisfying concordance. 

The transmission factors were obtained as follows: in 
the scalar case~($\sw=0$), we rewrote the radial equation into a Schr{\"o}dinger-like form and numerically integrated it using the scheme detailed in Ref.~\cite{Caio:2013}; in the electromagnetic~($\sw=-1$) and gravitational~($\sw=-2$) cases, we rewrote the radial Teukolsky equation using the \textit{Detweiler}~\cite{detweiler1976equations} and \textit{Sasaki-Nakamura}~\cite{sasaki1982gravitational} transformations, respectively. We numerically integrated the Detweiler and Sasaki-Nakamura equations from $r=r_{h}$ to $r = r_{\infty}$, where $r_h \sim 1.001 r_+$ and $r_\infty \sim 10^3 r_+$ are within the near-horizon and the far-field regimes, respectively. At $r=r_{\infty}$, we extract the values of the ingoing and outgoing coefficients via (\ref{eq:ingoing_sol}) and compute the transmission factors via (\ref{eq:trans}). To assure the reliability of our results, we have checked them using independent codes \cite{Dolan:2008kf}.    

\begin{figure}
\includegraphics[width=\columnwidth]{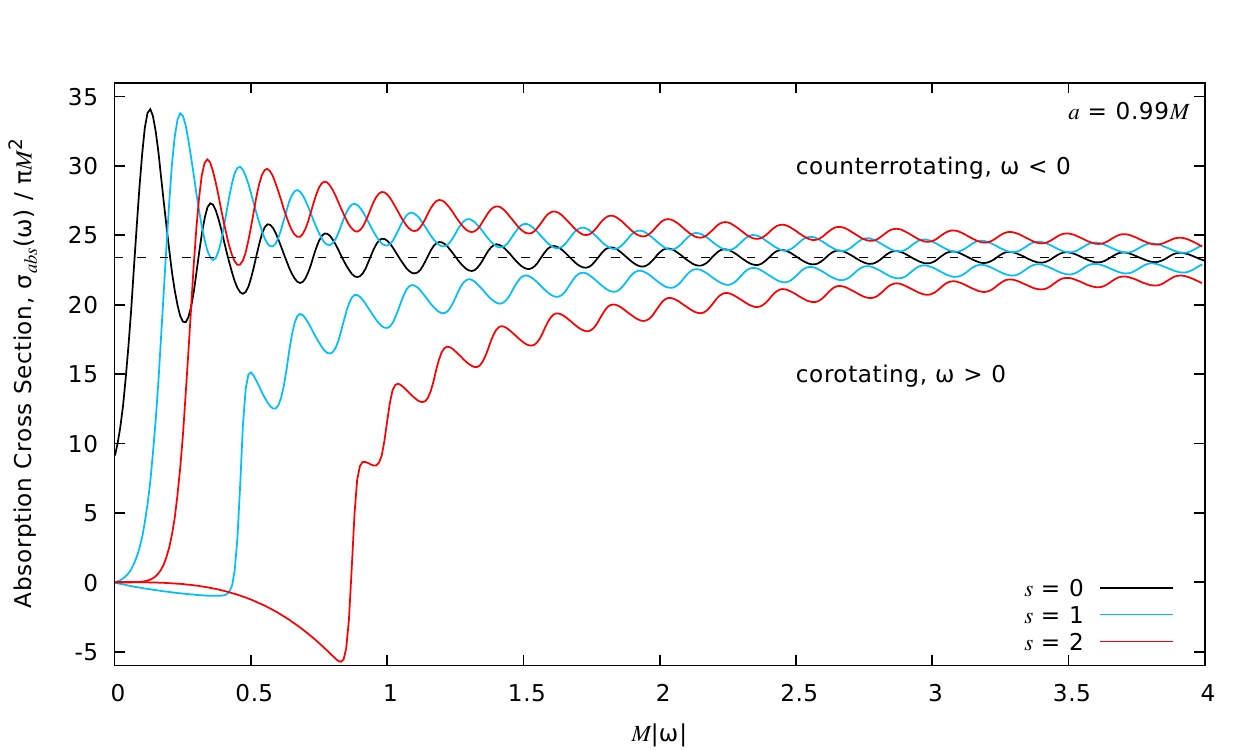}%%
\caption{The absorption cross section $\sigabs$ for massless bosonic fields incident on a rapidly-rotating Kerr BH~($a=0.99M$, $\gamma = 0$). For circularly-polarized fields ($s > 0$), the co-rotating~($\omega>0$) and counter-rotating ($\omega<0$) helicities~ are absorbed differently, due a coupling between the field helicity and the BH rotation.} 
\label{fig:allspins_a099_pol_abs}
\end{figure}

%%%%%%%%%%%%%%%%%%%%%%%%%%%%%%%%%%%%%%%%%%%%%%%%%%%%%%%%%%%%%%
{\it Numerical results.}---
Figure \ref{fig:allspins_a099_pol_abs} shows the absorption cross section $\sigabs$ for planar waves in all massless bosonic fields ($s=0$, $1$ and $2$) impinging upon a rapidly-rotating Kerr BH~($a^\ast=0.99$) parallel to the rotation axis ($\gamma = 0$). 
At long wavelengths, the incident wave stimulates superradiant emission from the black hole \cite{Rosa:2016bli}, with transmission turning negative for modes satisfying $\omega (\omega - m \Omega_h) < 0$.  For on-axis incidence $\gamma = 0$, only the $m = -\sw$ modes contribute to the mode sum (\ref{eq:absorption_cs}). Thus, $\sigabs$ is negative for polarized fields ($s >0$), but not for the scalar field ($s=0$). The superradiant effect occurs principally in the $l = m = -\sw$ mode, and is much stronger for gravitational waves than for electromagnetic waves. 

The absorption cross section for the co- and counter-rotating helicities are quite distinct, with the latter ($\omega < 0$) more strongly absorbed than the former ($\omega > 0$). This is a clear manifestation of the spin-helicity effect for electromagnetic and gravitational waves. In the limit $M |\omega| \rightarrow \infty$, the difference falls off at $O(M|\omega|)^{-1}$ and $\sigabs$ approaches the geodesic capture cross section $\sigraygeo$. We now attempt to quantify this effect.

{\it High frequency model.}---
Figure \ref{fig:allspins_a099_pol_abs} exhibits regular oscillations in $\sigabs(\omega)$ arising from successive $l$ modes in Eq.~(\ref{eq:absorption_cs}). For scalar fields it was previously shown \cite{Decanini:2011xi, Caio:2013} that such oscillations are linked to the Regge pole spectrum of the black hole, whose asymptotic properties are set by the angular frequency $\Omega_c$ and Lyapunov exponent $\Lambda_c$ of the circular photon orbits of the spacetime. At high frequencies for $\gamma = 0$, $\sigabs$ is well described by the \emph{sinc approximation} \cite{Sanchez:1977si,Decanini:2011xi, Caio:2013},
\beq
\sigabs \approx \sigma_{\text{sinc}} \equiv \calC_{s} + \eps \calA_{s} \sin\left(\calB_s / \eps \right) , \label{eq:sinc}
\eeq 
where $\eps \equiv (M |\omega|)^{-1}$. For massless scalar fields ($s=0$), it was shown in Ref.~\cite{Caio:2013} that Eq.~(\ref{eq:sinc}) applies with
\beq
 \calA_0 = -\frac{4 \pi \Lambda_c e^{-\pi \Lambda_c / \Omega_c}}{\Omega_c^2} , \quad  
\calB_0 = \frac{2\pi}{M \Omega_c}, \label{eq:geopar}
\eeq
and $\calC_{0} = \sigraygeo = \pi b_c^2$. Sample values for $b_c$, $\Omega_c$ and $\Lambda_c$ are given in Table \ref{tbl:geo}. The method for obtaining these values is covered in Ref.~\cite{Caio:2013}.

\begin{table}
\begin{tabular}{c | ccccc}
 $a^\ast$ & $0$ & $0.5$ & $0.8$ & $0.99$ & $1$ \\
 \hline
 $b_c/M$ & $\sqrt{27}$ & 5.1205 & 4.9849 & 4.8383 & 4.8284  \\
 $\Omega_c M$ & $\frac{1}{\sqrt{27}}$ & 0.1958 & 0.2019 & 0.2089 & 0.2094 \\
 $\Lambda_c M$ & $\frac{1}{\sqrt{27}}$ & 0.1884 & 0.1788 & 0.1633 & 0.1620
 \end{tabular}
\caption{The impact parameter $b_c$, orbital frequency $\Omega_c$ and Lyapunov exponent $\Lambda_c$ for circular polar null geodesics, to four decimal places. See Eq.~(\ref{eq:geopar}).}
\label{tbl:geo}
\end{table}

\begin{figure}
%\begin{figure*}
\includegraphics[width=\columnwidth]{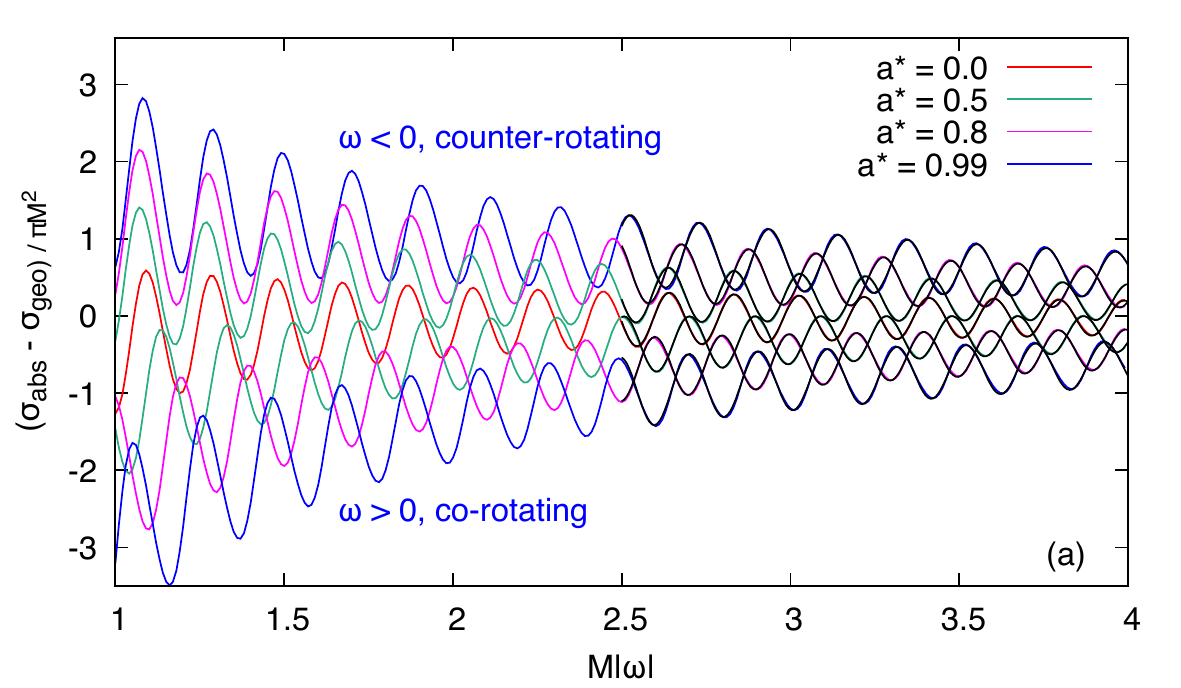}
\includegraphics[width=\columnwidth]{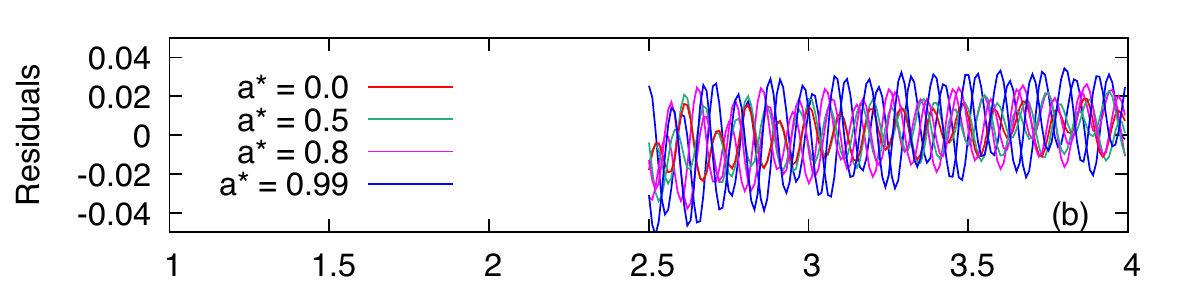}
\includegraphics[width=\columnwidth]{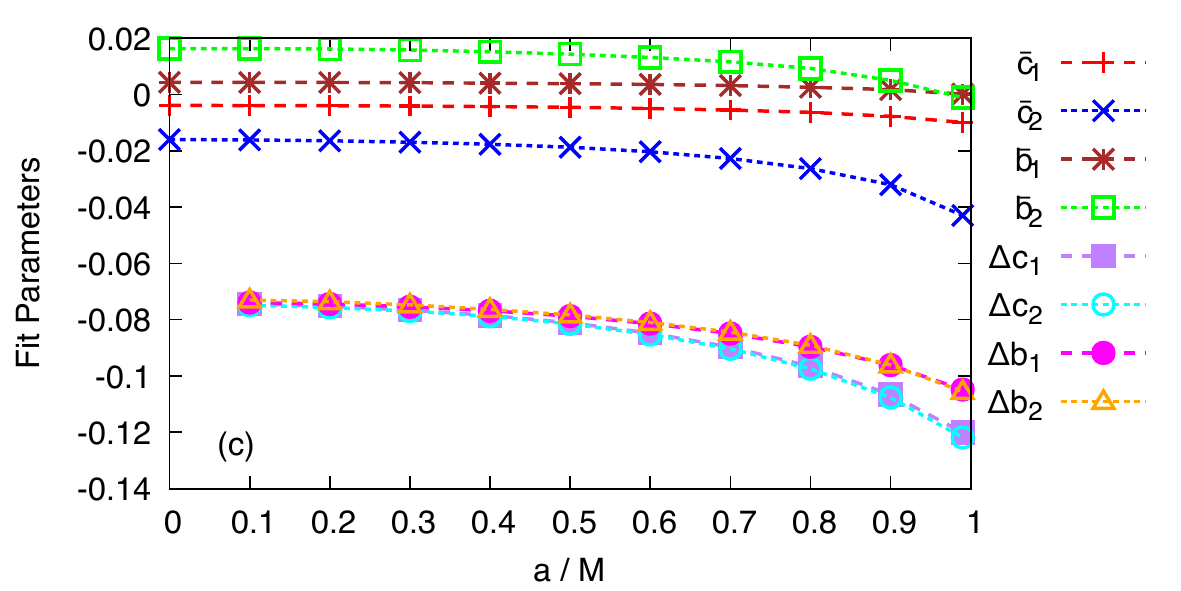}
\caption{(a) Fitting the sinc approximation model (\ref{eq:sinc})--(\ref{eq:model}) to numerical data for $a^\ast \in \{0, 0.5, 0.8, 0.99\}$ across the domain $M|\omega| \in [2.5, 4]$. (b) The residuals of the fit, $|\sigabs -  \sigma_s^{\text{ray}}| / \pi M^2$. (c) The best-fit values for the parameters $\{\bar{c}_s, \bar{b}_s, \Delta{c}_s, \Delta{b}_s \}$ in Eq.~(\ref{eq:model}).}
\label{fig:fit}
\end{figure}

For $s > 0$, we now propose an extended model which includes terms at $O(\eps)$:
\begin{subequations}
\begin{align}
\calB_{s>0} &= \calB_0 \left[1 + \eps \left(\bar{b}_{s} \pm s \, a^\ast \Delta b_{s}  \right) + O(\eps^2) \right],
\label{eq:modelb} \\
\calC_{s>0} &= \calC_0 \left[1 + \eps \left(\bar{c}_{s} \pm s \, a^\ast \Delta c_{s}  \right) + O(\eps^2) \right],
\label{eq:modelc} 
\end{align}
\label{eq:model} 
\end{subequations}
and $\calA_{s>0} = \calA_{0}$. The coefficients $\Delta b_s$ and $\Delta c_s$ encapsulate the effect of the spin-helicity interaction, with $+$ in Eq.~(\ref{eq:model}) for the co-rotating helicity, and $-$ for the counter-rotating helicity. 
To find the coefficients we fitted the model to our numerical data $\sigabs$ across the domain $M|\omega| \in [2.5,4]$ for $0 \le a^\ast \le 0.99$. Figure \ref{fig:fit} shows that the model (\ref{eq:sinc})--(\ref{eq:model}) fits the data well across the domain in $\omega$. 

%\end{figure*}

We may draw several inferences from the best-fit parameter values shown in Fig.~\ref{fig:fit}(c). First, that $\Delta b_1 = \Delta b_2$ and $\Delta c_1 = \Delta c_2$ to within the fitting error. This implies that the spin-helicity effect for gravitational waves is twice as large as for electromagnetic waves, as expected. Second, that $\Delta c_s \rightarrow \Delta b_s$ as $a^\ast \rightarrow 0$, which was not anticipated \emph{a priori}. Third, that $\calB_0 \Delta b_s  s a^\ast$, the spin-helicity part of the phase term in the sinc approximation (\ref{eq:sinc}), varies monotonically from $0$ in the Schwarzschild case up to approximately $s \pi$ in the extremal limit ($a \rightarrow M$). Evidence of this phase shift can be seen in Fig.~\ref{fig:fit}(a). 

%%%%%%%%%%%%%%%%%%%%%%%%%%%%%%%%%%%%%%%%%%%%%%%%%%%
{\it Final remarks.}---We have calculated the absorption cross section for scalar, 
electromagnetic, and gravitational massless plane waves impinging upon a Kerr BH along its rotation 
axis. For the first time, we have presented a unified picture of the absorption spectrum for all 
the bosonic fields. We showed that superradiance can overcome absorption, leading to $\sigabs < 0$ at low frequencies for co-rotating circular polarizations; and that counter-rotating polarizations are more heavily absorbed in general. We have proposed and tested an extended version of the sinc approximation, to encapsulate the spin-helicity effect at short wavelengths, where its effect falls off with $\lambda / M$. 

An open question is whether the spin-helicity effect shown here can be quantitatively described using \emph{spinoptics} \cite{Frolov:2011mh, Frolov:2012zn, Yoo:2012vv}. That is, can a modified geometric-optics approximation, incorporating next-to-leading order helicity-dependent corrections in the eikonal equations, successfully reproduce the $O(\eps)$ terms in Eqs.~(\ref{eq:model})? Future work in this direction could prove illuminating. 

%%%%%%%%%%%%%%%%%%%%%%%%%%%%%%%%%%%%%%%%%%%%%%%%%%%%%%%
\begin{acknowledgments}
%%%%%%%%%%%%%%%%%%%%%%%%%%%%%%
The authors would like to thank Conselho Nacional de Desenvolvimento Cient\'ifico e Tecnol\'ogico (CNPq) and Coordena\c{c}\~ao de Aperfei\c{c}oamento de Pessoal de N\'ivel Superior (CAPES), in Brazil, for partial financial support. 
S.D.~acknowledges financial support from the Engineering and Physical Sciences Research Council (EPSRC) under Grant No.~EP/M025802/1 and from the Science and Technology Facilities Council (STFC) under Grant No.~ST/L000520/1.
\end{acknowledgments}
%
%%%%%%%%%%%%%%%%%%%%%%%%%%%%%%%%%%%%%%%%%%%%%%%%%%%%%%

\bibliographystyle{apsrev4-1}
\bibliography{refs}

\begin{thebibliography}{31}%
\makeatletter
\providecommand \@ifxundefined [1]{%
 \@ifx{#1\undefined}
}%
\providecommand \@ifnum [1]{%
 \ifnum #1\expandafter \@firstoftwo
 \else \expandafter \@secondoftwo
 \fi
}%
\providecommand \@ifx [1]{%
 \ifx #1\expandafter \@firstoftwo
 \else \expandafter \@secondoftwo
 \fi
}%
\providecommand \natexlab [1]{#1}%
\providecommand \enquote  [1]{``#1''}%
\providecommand \bibnamefont  [1]{#1}%
\providecommand \bibfnamefont [1]{#1}%
\providecommand \citenamefont [1]{#1}%
\providecommand \href@noop [0]{\@secondoftwo}%
\providecommand \href [0]{\begingroup \@sanitize@url \@href}%
\providecommand \@href[1]{\@@startlink{#1}\@@href}%
\providecommand \@@href[1]{\endgroup#1\@@endlink}%
\providecommand \@sanitize@url [0]{\catcode `\\12\catcode `\$12\catcode
  `\&12\catcode `\#12\catcode `\^12\catcode `\_12\catcode `\%12\relax}%
\providecommand \@@startlink[1]{}%
\providecommand \@@endlink[0]{}%
\providecommand \url  [0]{\begingroup\@sanitize@url \@url }%
\providecommand \@url [1]{\endgroup\@href {#1}{\urlprefix }}%
\providecommand \urlprefix  [0]{URL }%
\providecommand \Eprint [0]{\href }%
\providecommand \doibase [0]{http://dx.doi.org/}%
\providecommand \selectlanguage [0]{\@gobble}%
\providecommand \bibinfo  [0]{\@secondoftwo}%
\providecommand \bibfield  [0]{\@secondoftwo}%
\providecommand \translation [1]{[#1]}%
\providecommand \BibitemOpen [0]{}%
\providecommand \bibitemStop [0]{}%
\providecommand \bibitemNoStop [0]{.\EOS\space}%
\providecommand \EOS [0]{\spacefactor3000\relax}%
\providecommand \BibitemShut  [1]{\csname bibitem#1\endcsname}%
\let\auto@bib@innerbib\@empty
%</preamble>
\bibitem [{\citenamefont {Thorne}\ \emph {et~al.}(1986)\citenamefont {Thorne},
  \citenamefont {Price},\ and\ \citenamefont {MacDonald}}]{Thorne:1986}%
  \BibitemOpen
  \bibfield  {author} {\bibinfo {author} {\bibfnamefont {K.~S.}\ \bibnamefont
  {Thorne}}, \bibinfo {author} {\bibfnamefont {R.~H.}\ \bibnamefont {Price}}, \
  and\ \bibinfo {author} {\bibfnamefont {D.~A.}\ \bibnamefont {MacDonald}},\
  }\href@noop {} {\emph {\bibinfo {title} {Black holes: the membrane
  paradigm}}}\ (\bibinfo  {publisher} {Yale University Press},\ \bibinfo {year}
  {1986})\BibitemShut {NoStop}%
\bibitem [{\citenamefont {Frolov}\ and\ \citenamefont
  {Novikov}(1998)}]{Frolov:1998wf}%
  \BibitemOpen
  \bibinfo {editor} {\bibfnamefont {V.~P.}\ \bibnamefont {Frolov}}\ and\
  \bibinfo {editor} {\bibfnamefont {I.~D.}\ \bibnamefont {Novikov}},\ eds.,\
  \href@noop {} {\emph {\bibinfo {title} {{Black hole physics: Basic concepts
  and new developments}}}}\ (\bibinfo {year} {1998})\BibitemShut {NoStop}%
%%CITATION = INSPIRE-486267;%%
\bibitem [{\citenamefont {Wald}(2001)}]{Wald:1999vt}%
  \BibitemOpen
  \bibfield  {author} {\bibinfo {author} {\bibfnamefont {R.~M.}\ \bibnamefont
  {Wald}},\ }\href {\doibase 10.12942/lrr-2001-6} {\bibfield  {journal}
  {\bibinfo  {journal} {Living Rev. Rel.}\ }\textbf {\bibinfo {volume} {4}},\
  \bibinfo {pages} {6} (\bibinfo {year} {2001})},\ \Eprint
  {http://arxiv.org/abs/gr-qc/9912119} {arXiv:gr-qc/9912119 [gr-qc]}
  \BibitemShut {NoStop}%
%%CITATION = GR-QC/9912119;%%
\bibitem [{\citenamefont {Abbott}\ \emph
  {et~al.}(2016{\natexlab{a}})\citenamefont {Abbott} \emph
  {et~al.}}]{Abbott:2016blz}%
  \BibitemOpen
  \bibfield  {author} {\bibinfo {author} {\bibfnamefont {B.~P.}\ \bibnamefont
  {Abbott}} \emph {et~al.} (\bibinfo {collaboration} {Virgo, LIGO
  Scientific}),\ }\href {\doibase 10.1103/PhysRevLett.116.061102} {\bibfield
  {journal} {\bibinfo  {journal} {Phys. Rev. Lett.}\ }\textbf {\bibinfo
  {volume} {116}},\ \bibinfo {pages} {061102} (\bibinfo {year}
  {2016}{\natexlab{a}})},\ \Eprint {http://arxiv.org/abs/1602.03837}
  {arXiv:1602.03837 [gr-qc]} \BibitemShut {NoStop}%
%%CITATION = ARXIV:1602.03837;%%
\bibitem [{\citenamefont {Abbott}\ \emph
  {et~al.}(2016{\natexlab{b}})\citenamefont {Abbott} \emph
  {et~al.}}]{TheLIGOScientific:2016pea}%
  \BibitemOpen
  \bibfield  {author} {\bibinfo {author} {\bibfnamefont {B.~P.}\ \bibnamefont
  {Abbott}} \emph {et~al.} (\bibinfo {collaboration} {Virgo, LIGO
  Scientific}),\ }\href {\doibase 10.1103/PhysRevX.6.041015} {\bibfield
  {journal} {\bibinfo  {journal} {Phys. Rev.}\ }\textbf {\bibinfo {volume}
  {X6}},\ \bibinfo {pages} {041015} (\bibinfo {year} {2016}{\natexlab{b}})},\
  \Eprint {http://arxiv.org/abs/1606.04856} {arXiv:1606.04856 [gr-qc]}
  \BibitemShut {NoStop}%
%%CITATION = ARXIV:1606.04856;%%
\bibitem [{\citenamefont {Ricarte}\ and\ \citenamefont
  {Dexter}(2015)}]{Ricarte:2014nca}%
  \BibitemOpen
  \bibfield  {author} {\bibinfo {author} {\bibfnamefont {A.}~\bibnamefont
  {Ricarte}}\ and\ \bibinfo {author} {\bibfnamefont {J.}~\bibnamefont
  {Dexter}},\ }\href {\doibase 10.1093/mnras/stu2128} {\bibfield  {journal}
  {\bibinfo  {journal} {Mon. Not. Roy. Astron. Soc.}\ }\textbf {\bibinfo
  {volume} {446}},\ \bibinfo {pages} {1973} (\bibinfo {year} {2015})},\ \Eprint
  {http://arxiv.org/abs/1410.2899} {arXiv:1410.2899 [astro-ph.HE]} \BibitemShut
  {NoStop}%
%%CITATION = ARXIV:1410.2899;%%
\bibitem [{\citenamefont {Castelvecchi}(2017)}]{castelvecchi2017hunt}%
  \BibitemOpen
  \bibfield  {author} {\bibinfo {author} {\bibfnamefont {D.}~\bibnamefont
  {Castelvecchi}},\ }\href@noop {} {\bibfield  {journal} {\bibinfo  {journal}
  {Nature}\ }\textbf {\bibinfo {volume} {543}},\ \bibinfo {pages} {478}
  (\bibinfo {year} {2017})}\BibitemShut {NoStop}%
\bibitem [{\citenamefont {Falcke}\ \emph {et~al.}(2000)\citenamefont {Falcke},
  \citenamefont {Melia},\ and\ \citenamefont {Agol}}]{Falcke:1999pj}%
  \BibitemOpen
  \bibfield  {author} {\bibinfo {author} {\bibfnamefont {H.}~\bibnamefont
  {Falcke}}, \bibinfo {author} {\bibfnamefont {F.}~\bibnamefont {Melia}}, \
  and\ \bibinfo {author} {\bibfnamefont {E.}~\bibnamefont {Agol}},\ }\href
  {\doibase 10.1086/312423} {\bibfield  {journal} {\bibinfo  {journal}
  {Astrophys. J.}\ }\textbf {\bibinfo {volume} {528}},\ \bibinfo {pages} {L13}
  (\bibinfo {year} {2000})},\ \Eprint {http://arxiv.org/abs/astro-ph/9912263}
  {arXiv:astro-ph/9912263 [astro-ph]} \BibitemShut {NoStop}%
%%CITATION = ASTRO-PH/9912263;%%
\bibitem [{\citenamefont {Lu}\ \emph {et~al.}(2014)\citenamefont {Lu},
  \citenamefont {Broderick}, \citenamefont {Baron}, \citenamefont {Monnier},
  \citenamefont {Fish}, \citenamefont {Doeleman},\ and\ \citenamefont
  {Pankratius}}]{Lu:2014zja}%
  \BibitemOpen
  \bibfield  {author} {\bibinfo {author} {\bibfnamefont {R.-S.}\ \bibnamefont
  {Lu}}, \bibinfo {author} {\bibfnamefont {A.~E.}\ \bibnamefont {Broderick}},
  \bibinfo {author} {\bibfnamefont {F.}~\bibnamefont {Baron}}, \bibinfo
  {author} {\bibfnamefont {J.~D.}\ \bibnamefont {Monnier}}, \bibinfo {author}
  {\bibfnamefont {V.~L.}\ \bibnamefont {Fish}}, \bibinfo {author}
  {\bibfnamefont {S.~S.}\ \bibnamefont {Doeleman}}, \ and\ \bibinfo {author}
  {\bibfnamefont {V.}~\bibnamefont {Pankratius}},\ }\href {\doibase
  10.1088/0004-637X/788/2/120} {\bibfield  {journal} {\bibinfo  {journal}
  {Astrophys. J.}\ }\textbf {\bibinfo {volume} {788}},\ \bibinfo {pages} {120}
  (\bibinfo {year} {2014})},\ \Eprint {http://arxiv.org/abs/1404.7095}
  {arXiv:1404.7095 [astro-ph.IM]} \BibitemShut {NoStop}%
%%CITATION = ARXIV:1404.7095;%%
\bibitem [{\citenamefont {Johannsen}(2016)}]{Johannsen:2016uoh}%
  \BibitemOpen
  \bibfield  {author} {\bibinfo {author} {\bibfnamefont {T.}~\bibnamefont
  {Johannsen}},\ }\href {\doibase 10.1088/0264-9381/33/12/124001} {\bibfield
  {journal} {\bibinfo  {journal} {Class. Quant. Grav.}\ }\textbf {\bibinfo
  {volume} {33}},\ \bibinfo {pages} {124001} (\bibinfo {year} {2016})},\
  \Eprint {http://arxiv.org/abs/1602.07694} {arXiv:1602.07694 [astro-ph.HE]}
  \BibitemShut {NoStop}%
%%CITATION = ARXIV:1602.07694;%%
\bibitem [{\citenamefont {Akiyama}\ \emph {et~al.}(2017)\citenamefont {Akiyama}
  \emph {et~al.}}]{Akiyama:2017rcc}%
  \BibitemOpen
  \bibfield  {author} {\bibinfo {author} {\bibfnamefont {K.}~\bibnamefont
  {Akiyama}} \emph {et~al.},\ }\href {\doibase 10.3847/1538-4357/aa6305}
  {\bibfield  {journal} {\bibinfo  {journal} {Astrophys. J.}\ }\textbf
  {\bibinfo {volume} {838}},\ \bibinfo {pages} {1} (\bibinfo {year} {2017})},\
  \Eprint {http://arxiv.org/abs/1702.07361} {arXiv:1702.07361 [astro-ph.IM]}
  \BibitemShut {NoStop}%
%%CITATION = ARXIV:1702.07361;%%
\bibitem [{\citenamefont {Glampedakis}\ and\ \citenamefont
  {Andersson}(2001)}]{Glampedakis:2001cx}%
  \BibitemOpen
  \bibfield  {author} {\bibinfo {author} {\bibfnamefont {K.}~\bibnamefont
  {Glampedakis}}\ and\ \bibinfo {author} {\bibfnamefont {N.}~\bibnamefont
  {Andersson}},\ }\href {\doibase 10.1088/0264-9381/18/10/309} {\bibfield
  {journal} {\bibinfo  {journal} {Classical Quantum Gravity}\ }\textbf
  {\bibinfo {volume} {18}},\ \bibinfo {pages} {1939} (\bibinfo {year}
  {2001})},\ \Eprint {http://arxiv.org/abs/gr-qc/0102100} {arXiv:gr-qc/0102100
  [gr-qc]} \BibitemShut {NoStop}%
%%CITATION = GR-QC/0102100;%%
\bibitem [{\citenamefont {Dolan}(2008)}]{Dolan:2008kf}%
  \BibitemOpen
  \bibfield  {author} {\bibinfo {author} {\bibfnamefont {S.~R.}\ \bibnamefont
  {Dolan}},\ }\href {\doibase 10.1088/0264-9381/25/23/235002} {\bibfield
  {journal} {\bibinfo  {journal} {Classical Quantum Gravity}\ }\textbf
  {\bibinfo {volume} {25}},\ \bibinfo {pages} {235002} (\bibinfo {year}
  {2008})},\ \Eprint {http://arxiv.org/abs/0801.3805} {arXiv:0801.3805 [gr-qc]}
  \BibitemShut {NoStop}%
%%CITATION = ARXIV:0801.3805;%%
\bibitem [{\citenamefont {Macedo}\ \emph {et~al.}(2013)\citenamefont {Macedo},
  \citenamefont {Leite}, \citenamefont {Oliveira}, \citenamefont {Dolan},\ and\
  \citenamefont {Crispino}}]{Caio:2013}%
  \BibitemOpen
  \bibfield  {author} {\bibinfo {author} {\bibfnamefont {C.~F.~B.}\
  \bibnamefont {Macedo}}, \bibinfo {author} {\bibfnamefont {L.~C.~S.}\
  \bibnamefont {Leite}}, \bibinfo {author} {\bibfnamefont {E.~S.}\ \bibnamefont
  {Oliveira}}, \bibinfo {author} {\bibfnamefont {S.~R.}\ \bibnamefont {Dolan}},
  \ and\ \bibinfo {author} {\bibfnamefont {L.~C.~B.}\ \bibnamefont
  {Crispino}},\ }\href {\doibase 10.1103/PhysRevD.88.064033} {\bibfield
  {journal} {\bibinfo  {journal} {Phys. Rev. D}\ }\textbf {\bibinfo {volume}
  {88}},\ \bibinfo {pages} {064033} (\bibinfo {year} {2013})}\BibitemShut
  {NoStop}%
\bibitem [{\citenamefont {Abbott}\ \emph
  {et~al.}(2016{\natexlab{c}})\citenamefont {Abbott} \emph
  {et~al.}}]{Abbott:2016bqf}%
  \BibitemOpen
  \bibfield  {author} {\bibinfo {author} {\bibfnamefont {B.~P.}\ \bibnamefont
  {Abbott}} \emph {et~al.} (\bibinfo {collaboration} {Virgo, LIGO
  Scientific}),\ }\href {\doibase 10.1002/andp.201600209} {\bibfield  {journal}
  {\bibinfo  {journal} {Annalen Phys.}\ } (\bibinfo {year}
  {2016}{\natexlab{c}}),\ 10.1002/andp.201600209},\ \bibinfo {note} {[Annalen
  Phys.529,0209(2017)]},\ \Eprint {http://arxiv.org/abs/1608.01940}
  {arXiv:1608.01940 [gr-qc]} \BibitemShut {NoStop}%
%%CITATION = ARXIV:1608.01940;%%
\bibitem [{\citenamefont {Synge}(1966)}]{Synge:1966}%
  \BibitemOpen
  \bibfield  {author} {\bibinfo {author} {\bibfnamefont {J.~L.}\ \bibnamefont
  {Synge}},\ }\href@noop {} {\bibfield  {journal} {\bibinfo  {journal} {Monthly
  Notices of the Royal Astronomical Society}\ }\textbf {\bibinfo {volume}
  {131}},\ \bibinfo {pages} {463} (\bibinfo {year} {1966})}\BibitemShut
  {NoStop}%
\bibitem [{\citenamefont {Grenzebach}\ \emph {et~al.}(2014)\citenamefont
  {Grenzebach}, \citenamefont {Perlick},\ and\ \citenamefont
  {Lämmerzahl}}]{Grenzebach:2014fha}%
  \BibitemOpen
  \bibfield  {author} {\bibinfo {author} {\bibfnamefont {A.}~\bibnamefont
  {Grenzebach}}, \bibinfo {author} {\bibfnamefont {V.}~\bibnamefont {Perlick}},
  \ and\ \bibinfo {author} {\bibfnamefont {C.}~\bibnamefont {Lämmerzahl}},\
  }\href {\doibase 10.1103/PhysRevD.89.124004} {\bibfield  {journal} {\bibinfo
  {journal} {Phys. Rev.}\ }\textbf {\bibinfo {volume} {D89}},\ \bibinfo {pages}
  {124004} (\bibinfo {year} {2014})},\ \Eprint {http://arxiv.org/abs/1403.5234}
  {arXiv:1403.5234 [gr-qc]} \BibitemShut {NoStop}%
%%CITATION = ARXIV:1403.5234;%%
\bibitem [{\citenamefont {Brito}\ \emph {et~al.}(2015)\citenamefont {Brito},
  \citenamefont {Cardoso},\ and\ \citenamefont {Pani}}]{Brito:2015oca}%
  \BibitemOpen
  \bibfield  {author} {\bibinfo {author} {\bibfnamefont {R.}~\bibnamefont
  {Brito}}, \bibinfo {author} {\bibfnamefont {V.}~\bibnamefont {Cardoso}}, \
  and\ \bibinfo {author} {\bibfnamefont {P.}~\bibnamefont {Pani}},\ }\href
  {\doibase 10.1007/978-3-319-19000-6} {\bibfield  {journal} {\bibinfo
  {journal} {Lect. Notes Phys.}\ }\textbf {\bibinfo {volume} {906}},\ \bibinfo
  {pages} {pp.1} (\bibinfo {year} {2015})},\ \Eprint
  {http://arxiv.org/abs/1501.06570} {arXiv:1501.06570 [gr-qc]} \BibitemShut
  {NoStop}%
%%CITATION = ARXIV:1501.06570;%%
\bibitem [{\citenamefont {Mashhoon}(1974)}]{Mashhoon:1974cq}%
  \BibitemOpen
  \bibfield  {author} {\bibinfo {author} {\bibfnamefont {B.}~\bibnamefont
  {Mashhoon}},\ }\href {\doibase 10.1103/PhysRevD.10.1059} {\bibfield
  {journal} {\bibinfo  {journal} {Phys. Rev.}\ }\textbf {\bibinfo {volume}
  {D10}},\ \bibinfo {pages} {1059} (\bibinfo {year} {1974})}\BibitemShut
  {NoStop}%
%%CITATION = PHRVA,D10,1059;%%
\bibitem [{\citenamefont {Frolov}\ and\ \citenamefont
  {Shoom}(2011)}]{Frolov:2011mh}%
  \BibitemOpen
  \bibfield  {author} {\bibinfo {author} {\bibfnamefont {V.~P.}\ \bibnamefont
  {Frolov}}\ and\ \bibinfo {author} {\bibfnamefont {A.~A.}\ \bibnamefont
  {Shoom}},\ }\href {\doibase 10.1103/PhysRevD.84.044026} {\bibfield  {journal}
  {\bibinfo  {journal} {Phys. Rev.}\ }\textbf {\bibinfo {volume} {D84}},\
  \bibinfo {pages} {044026} (\bibinfo {year} {2011})},\ \Eprint
  {http://arxiv.org/abs/1105.5629} {arXiv:1105.5629 [gr-qc]} \BibitemShut
  {NoStop}%
%%CITATION = ARXIV:1105.5629;%%
\bibitem [{\citenamefont {Frolov}\ and\ \citenamefont
  {Shoom}(2012)}]{Frolov:2012zn}%
  \BibitemOpen
  \bibfield  {author} {\bibinfo {author} {\bibfnamefont {V.~P.}\ \bibnamefont
  {Frolov}}\ and\ \bibinfo {author} {\bibfnamefont {A.~A.}\ \bibnamefont
  {Shoom}},\ }\href {\doibase 10.1103/PhysRevD.86.024010} {\bibfield  {journal}
  {\bibinfo  {journal} {Phys. Rev.}\ }\textbf {\bibinfo {volume} {D86}},\
  \bibinfo {pages} {024010} (\bibinfo {year} {2012})},\ \Eprint
  {http://arxiv.org/abs/1205.4479} {arXiv:1205.4479 [gr-qc]} \BibitemShut
  {NoStop}%
%%CITATION = ARXIV:1205.4479;%%
\bibitem [{\citenamefont {Yoo}(2012)}]{Yoo:2012vv}%
  \BibitemOpen
  \bibfield  {author} {\bibinfo {author} {\bibfnamefont {C.-M.}\ \bibnamefont
  {Yoo}},\ }\href {\doibase 10.1103/PhysRevD.86.084005} {\bibfield  {journal}
  {\bibinfo  {journal} {Phys. Rev.}\ }\textbf {\bibinfo {volume} {D86}},\
  \bibinfo {pages} {084005} (\bibinfo {year} {2012})},\ \Eprint
  {http://arxiv.org/abs/1207.6833} {arXiv:1207.6833 [gr-qc]} \BibitemShut
  {NoStop}%
%%CITATION = ARXIV:1207.6833;%%
\bibitem [{\citenamefont {Teukolsky}(1972)}]{teukolsky1972rotating}%
  \BibitemOpen
  \bibfield  {author} {\bibinfo {author} {\bibfnamefont {S.~A.}\ \bibnamefont
  {Teukolsky}},\ }\href@noop {} {\bibfield  {journal} {\bibinfo  {journal}
  {Physical Review Letters}\ }\textbf {\bibinfo {volume} {29}},\ \bibinfo
  {pages} {1114} (\bibinfo {year} {1972})}\BibitemShut {NoStop}%
\bibitem [{\citenamefont {{Futterman}}\ \emph {et~al.}(1988)\citenamefont
  {{Futterman}}, \citenamefont {{Handler}},\ and\ \citenamefont
  {{Matzner}}}]{Futterman:1988ni}%
  \BibitemOpen
  \bibfield  {author} {\bibinfo {author} {\bibfnamefont {J.~A.~H.}\
  \bibnamefont {{Futterman}}}, \bibinfo {author} {\bibfnamefont {F.~A.}\
  \bibnamefont {{Handler}}}, \ and\ \bibinfo {author} {\bibfnamefont {R.~A.}\
  \bibnamefont {{Matzner}}},\ }\href@noop {} {\emph {\bibinfo {title}
  {{Scattering from black holes}}}}\ (\bibinfo  {publisher} {Cambridge
  University Press},\ \bibinfo {year} {1988})\BibitemShut {NoStop}%
\bibitem [{\citenamefont {Cook}\ and\ \citenamefont {Zalutskiy}(2014)}]{Cook}%
  \BibitemOpen
  \bibfield  {author} {\bibinfo {author} {\bibfnamefont {G.~B.}\ \bibnamefont
  {Cook}}\ and\ \bibinfo {author} {\bibfnamefont {M.}~\bibnamefont
  {Zalutskiy}},\ }\href@noop {} {\bibfield  {journal} {\bibinfo  {journal}
  {Physical Review D}\ }\textbf {\bibinfo {volume} {90}},\ \bibinfo {pages}
  {124021} (\bibinfo {year} {2014})}\BibitemShut {NoStop}%
\bibitem [{\citenamefont {Berti}\ \emph {et~al.}(2006)\citenamefont {Berti},
  \citenamefont {Cardoso},\ and\ \citenamefont
  {Casals}}]{berti2006eigenvalues}%
  \BibitemOpen
  \bibfield  {author} {\bibinfo {author} {\bibfnamefont {E.}~\bibnamefont
  {Berti}}, \bibinfo {author} {\bibfnamefont {V.}~\bibnamefont {Cardoso}}, \
  and\ \bibinfo {author} {\bibfnamefont {M.}~\bibnamefont {Casals}},\
  }\href@noop {} {\bibfield  {journal} {\bibinfo  {journal} {Physical Review
  D}\ }\textbf {\bibinfo {volume} {73}},\ \bibinfo {pages} {024013} (\bibinfo
  {year} {2006})}\BibitemShut {NoStop}%
\bibitem [{\citenamefont {Detweiler}(1976)}]{detweiler1976equations}%
  \BibitemOpen
  \bibfield  {author} {\bibinfo {author} {\bibfnamefont {S.}~\bibnamefont
  {Detweiler}},\ }in\ \href@noop {} {\emph {\bibinfo {booktitle} {Proceedings
  of the Royal Society of London A: Mathematical, Physical and Engineering
  Sciences}}},\ Vol.\ \bibinfo {volume} {349}\ (\bibinfo {organization} {The
  Royal Society},\ \bibinfo {year} {1976})\ pp.\ \bibinfo {pages}
  {217--230}\BibitemShut {NoStop}%
\bibitem [{\citenamefont {Sasaki}\ and\ \citenamefont
  {Nakamura}(1982)}]{sasaki1982gravitational}%
  \BibitemOpen
  \bibfield  {author} {\bibinfo {author} {\bibfnamefont {M.}~\bibnamefont
  {Sasaki}}\ and\ \bibinfo {author} {\bibfnamefont {T.}~\bibnamefont
  {Nakamura}},\ }\href@noop {} {\bibfield  {journal} {\bibinfo  {journal}
  {Progress of Theoretical Physics}\ }\textbf {\bibinfo {volume} {67}},\
  \bibinfo {pages} {1788} (\bibinfo {year} {1982})}\BibitemShut {NoStop}%
\bibitem [{\citenamefont {Rosa}(2017)}]{Rosa:2016bli}%
  \BibitemOpen
  \bibfield  {author} {\bibinfo {author} {\bibfnamefont {J.~G.}\ \bibnamefont
  {Rosa}},\ }\href {\doibase 10.1103/PhysRevD.95.064017} {\bibfield  {journal}
  {\bibinfo  {journal} {Phys. Rev.}\ }\textbf {\bibinfo {volume} {D95}},\
  \bibinfo {pages} {064017} (\bibinfo {year} {2017})},\ \Eprint
  {http://arxiv.org/abs/1612.01826} {arXiv:1612.01826 [gr-qc]} \BibitemShut
  {NoStop}%
%%CITATION = ARXIV:1612.01826;%%
\bibitem [{\citenamefont {Decanini}\ \emph {et~al.}(2011)\citenamefont
  {Decanini}, \citenamefont {Esposito-Farese},\ and\ \citenamefont
  {Folacci}}]{Decanini:2011xi}%
  \BibitemOpen
  \bibfield  {author} {\bibinfo {author} {\bibfnamefont {Y.}~\bibnamefont
  {Decanini}}, \bibinfo {author} {\bibfnamefont {G.}~\bibnamefont
  {Esposito-Farese}}, \ and\ \bibinfo {author} {\bibfnamefont {A.}~\bibnamefont
  {Folacci}},\ }\href {\doibase 10.1103/PhysRevD.83.044032} {\bibfield
  {journal} {\bibinfo  {journal} {Phys. Rev. D}\ }\textbf {\bibinfo {volume}
  {83}},\ \bibinfo {pages} {044032} (\bibinfo {year} {2011})},\ \Eprint
  {http://arxiv.org/abs/1101.0781} {arXiv:1101.0781 [gr-qc]} \BibitemShut
  {NoStop}%
%%CITATION = ARXIV:1101.0781;%%
\bibitem [{\citenamefont {Sanchez}(1978)}]{Sanchez:1977si}%
  \BibitemOpen
  \bibfield  {author} {\bibinfo {author} {\bibfnamefont {N.~G.}\ \bibnamefont
  {Sanchez}},\ }\href {\doibase 10.1103/PhysRevD.18.1030} {\bibfield  {journal}
  {\bibinfo  {journal} {Phys. Rev.D}\ }\textbf {\bibinfo {volume} {18}},\
  \bibinfo {pages} {1030} (\bibinfo {year} {1978})}\BibitemShut {NoStop}%
%%CITATION = PHRVA,D18,1030;%%
\end{thebibliography}%
\end{document}